\documentclass[twocolumn,showpacs,preprintnumbers,amsmath,amssymb,pre]{revtex4}

\usepackage[dvips]{graphicx}
\usepackage{dcolumn}
\usepackage{amssymb}
\usepackage{bm}
\usepackage{latexsym}

\begin{document}

\preprint{APS}

\title{An $H$ theorem for contact forces in granular materials}

\author{Philip T. Metzger}
\email{Philip.T.Metzger@nasa.gov}
\affiliation{%
Granular Mechanics and Surface Systems Laboratory, John F. Kennedy Space Center, NASA\\
KT-D3, Kennedy Space Center, Florida  32899
}%

\date{\today}

\begin{abstract}
A maximum entropy theorem is developed and tested for granular contact forces.  Although it is idealized, describing two dimensional packings of round, rigid, frictionless, cohesionless disks with coordination number $Z=4$, it appears to describe a central part of the physics present in the more general cases.  The theorem does not make the strong claims of Edwards' hypothesis, nor does it rely upon Edwards' hypothesis at any point.  Instead, it begins solely from the physical assumption that closed loops of grains are unable to impose strong force correlations around the loop.  This statement is shown to be a generalization of Boltzmann's Assumption of Molecular Chaos (his \textit{stosszahlansatz}), allowing for the extra symmetries of granular stress propagation compared to the more limited symmetries of momentum propagation in a thermodynamic system.  The theorem that follows from this is similar to Boltzmann's $H$ theorem and is presented as an alternative to Edwards' hypothesis for explaining some granular phenomena.  It identifies a very interesting feature of granular packings:  if the generalized \textit{stosszahlansatz} is correct, then the bulk of homogeneous granular packings must satisfy a maximum entropy condition simply by virtue of being stable, without any exploration of phase space required.  This leads to an independent derivation of the contact force statistics, and these predictions have been compared to numerical simulation data in the isotropic case.  The good agreement implies that the generalized \textit{stosszahlansatz} is indeed accurate at least for the isotropic state of the idealized case studied here, and that it is the reductionist explanation for contact force statistics in this case.
\end{abstract}

\maketitle

\section{\label{sec:intro}Introduction}

Edwards hypothesized that every mechanically stable micro state of a powder is equally probable in a Gibbs-like ensemble, for cases where the powder is prepared in certain, repeatable ways \cite{edwards1}.  This hypothesis has been extended by others to the statistics of granular contact forces for both isostatic and hyperstatic cases.  Packings of round, rigid, frictionless disks or spheres are known to be isostatic \cite{isostatic}, meaning that there are exactly the same number of contact force unknowns in a granular packing as there are stability equations for the grains, so that the value of each contact force can be resolved by linear algebra operating solely upon the geometry of the inter-granular contact network (assuming that the overall stress of the packing has also been specified).  For isostatic cases, then, each valid micro state in an Edwards' ensemble corresponds to exactly one micro state of contact forces.  An ensemble of packings with a flat measure thus implies an ensemble of contact force networks with the same flat measure.  Hyperstatic packings, on the other hand, have more contact forces unknowns in the granular packing than there are stability equations for the individual grains.  Thus, the values of the forces throughout the packing are mechanically indeterminate.  A Gibbs-like ensemble method has been applied to the contact forces in such packings by selecting a single packing geometry that is held constant throughout the ensemble, assuming a flat measure over all of the valid contact force microstates \cite{snoeijer, tighe}.  All these ensembles, isostatic as well as hyperstatic, are similar to Gibbs' ensembles in thermal systems because the flat measure is assumed {\em a priori}.  The isostatic case was recently analyzed by extending Gibbs' {\em most probable distribution} method \cite{metzger2, metzger4}, the same method used in textbooks to derive the Maxwell-Boltzmann, Bose and Fermi distributions.  In those derivations, the most probable distribution is concerned with momenta $f(\bm p)$ or particle energies $f(E)$, but when extended to granular contact forces it predicts the most probable distribution of single grain states, $\rho(g)$, where $g$ is a set of variables that describe everything that can be known about an individual grain including all of its contact angles, forces and their correlations.  Subsequent discrete element modeling has validated the predictions of this theory for the special, isostatic case described above \cite{metzger4, metzger3}.  Thus, Edwards hypothesis is {\em sufficient} to derive granular contact force statistics for single grain states in this case.  

However, this paper will show that Edwards' hypothesis is not really {\em necessary} and that it is not really a part of the reductionist explanation for granular contact force statistics.  It makes a very strong claim about microstructural geometries being equally probable, and this is a much stronger claim than we need to derive granular contact force statistics.  Instead, the reductionist explanation is found in the nature of correlations in granular contact forces, which can be summarized in a statement that is a generalization of Boltzmann's Assumption of Molecular Chaos.  Boltzmann showed that the absence of pre-correlation between colliding gas molecules inexorably relaxes the gas to maximum entropy; this paper will show that the corresponding condition in granular contact forces inexorably requires the packing to exist (in its bulk) in a relaxed state of maximum contact force entropy.  This paper will follow Boltzmann's approach, developing a ``translation'' equation analogous to Boltzmann's transport equation (but adapted for static packings in which nothing is moving), with a maximum entropy proof similar to the $H$ theorem, to produce a new derivation of $\rho(g)$ without making any {\em a priori} assumptions of a flat measure.  This new theory is presented as an alternative to Edwards' hypothesis to provide a more physical basis for the statistical mechanics of some granular phenomena.

\section{\label{sec:AMC} The Granular Contact Force \textit{Stosszahlansatz}}

\subsection{``Translation'' in a Static Granular Packing}

\begin{figure}
\includegraphics[angle=0,width=1.0\columnwidth]{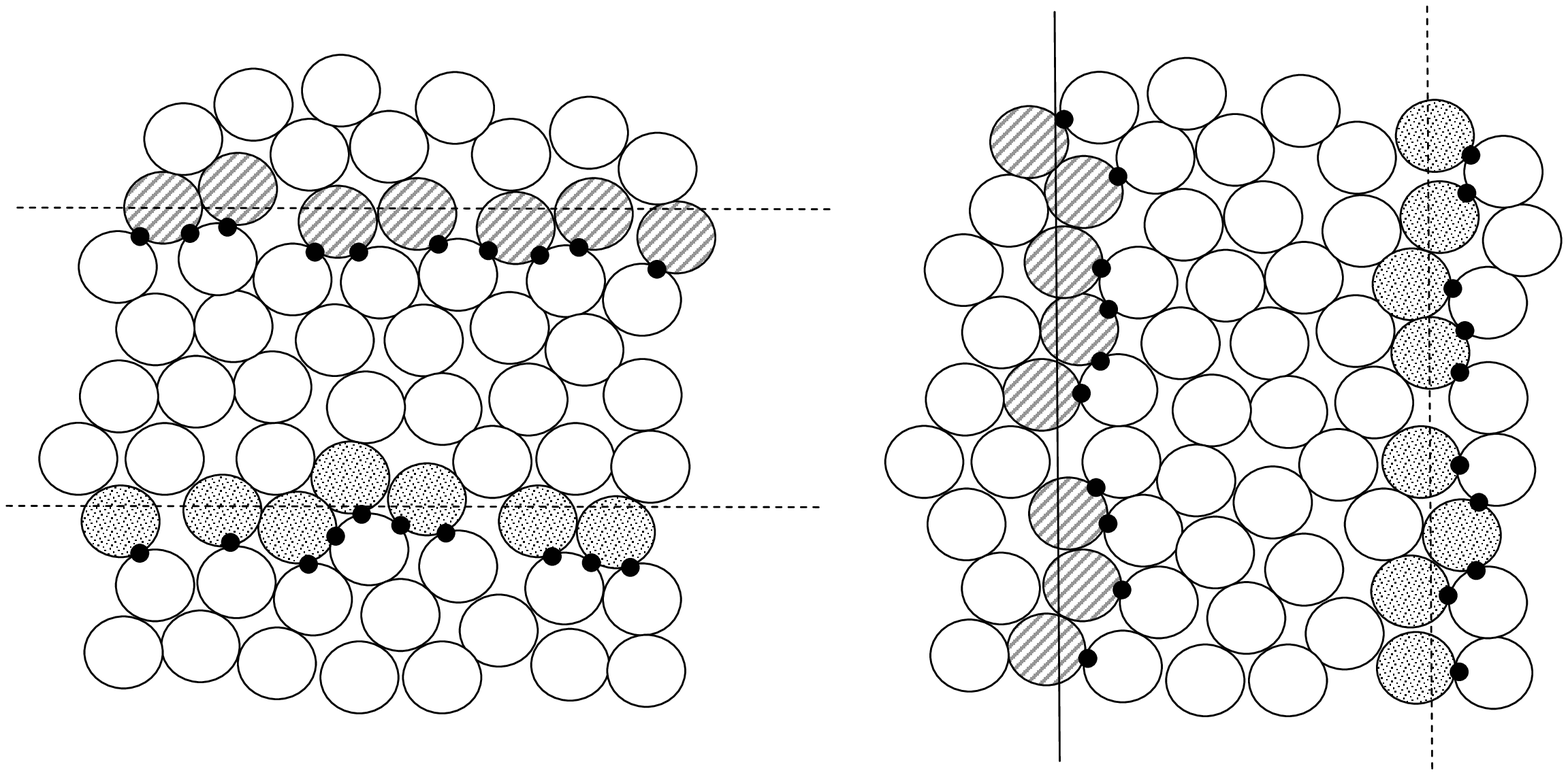}
\caption{\label{conserve} Illustration how sheets of force are conserved while translating through a 2D granular packing (as described in the text), so that a Boltzmann-like transport equation may be defined.}
\end{figure}

Boltzmann wrote a transport equation to track the evolution through time in the single particle distribution function $f(\bm{p})$, where $\bm{p}$ is momentum.  For {\em static} granular packings we must describe how the statistics evolve through space, not time.  We will therefore perform ``translation'' through successive cross-sectional ``layers'' of grains, as defined here.  Referring to Fig.~(\ref{conserve}),
we:  (1) draw a cross-sectional line (like any one of the straight lines in the figure), (2) select the set of grains intersected by that one line (shaded grains), and (3) include only those contacts on that set of grains (heavy dots) that (a) connect to other grains \textit{external} to the shaded set and (b) are located \textit{on one side} of the shaded set.  Next, we consider the force vectors on that set of contacts, including both the normal and tangential components of force if the grains are frictional.  We decompose these vectors into Cartesian components parallel and normal to the dashed line.  Summing all the components in the normal direction produces the ``total Cartesian load'' in that plane.  ``Translating'' refers to the continuous motion of the cross-sectional line that picks out the set of grains in a layer.  As the line translates across the packing, some grains are no longer intersected by the line and hence leave the layer, while some other grains become intersected by the line and hence join the layer.  In the absence of gravity, the total Cartesian load is conserved with respect to translating the straight line, as illustrated in the figure by the parallel layer of shaded grains and heavy dots.  If that were not so, then the grains between the two shaded layers would be accelerating.  

Because the container walls might also contribute forces parallel to the direction of translation, thus spoiling the conservation of the total force, we must consider special cases where this cannot occur.  There are at least two such cases, corresponding to microcanonical and canonical statistics.  If the container is 2D (for specificity), and if the side walls are flat, parallel and frictionless, then we will have two spatial axes along which to translate layers of grains subject to the conservation laws as shown in the left and right sides of Fig.~\ref{conserve}.  We cannot perform the translation in a diagonal direction in this finite container because then the length of layers will not be constant and neither will the total perpendicular force contained within them.  This translation in a finite container corresponds to microcanonical statistics because the system is closed and the force conservation is exact.  Alternatively, we can perform the translation in an extremely large packing (perhaps infinite) and consider only the set of grains selected by a relatively short line {\em segment} translating through the middle of the packing far from any boundaries.  In that case, the conservation law will not be exact because some force will be entering and exiting the segment at both of its ends throughout the translation.  However, the longer the segment of grains is, then the less significant those fluctuations become and the more closely it approximates an exact conservation law.  This corresponds to canonical statistics because the force contained in the finite line segment is in contact with an infinite reservoir of force at each of its ends.  In this canonical case the line segment may have any arbitrary orientation, not only the $x$ or $y$ directions, and it translates across the packing in a direction normal to its orientation.  In this paper it does not matter whether we consider the microcanonical or canonical case, nor in what direction we imagine the translation to occur.  The generality of the equations describing this translation contributes to the generality of the conclusions.

It should be noted that this translation \textit{ansatz} allows us to treat a static granular packing with methods borrowed from kinetic theory, even though the packing is completely static.  The role of the time dimension from kinetic theory is replaced by the translation through a spatial dimension.  The collisions of particles from kinetic theory are replaced by the meeting together of contact forces on the grains.  The force vectors on one hemisphere of a grain are analogous to gas particles going into a collision, and the forces on the opposite hemisphere are analogous to the same particles exiting from the collision later in time.  (The number of contact forces ``entering'' and ``exiting'' the opposite hemispheres of a grain in this fashion need not be conserved.)  Thus, the granular theory developed here is analagous to molecular kinetic theory, but it must be remembered that everything is completely static and that all contacts between grains are unchanging in time, not transitory like the collisions in kinetic theory.  This theory is not to be confused with the kinetic theory of granular gases, which deals with grains moving and colliding in time.

Although this force conservation applies to more general cases, the remainder of this paper deals only with round, rigid, frictionless, 2D grains.  Table~\ref{convention} defines the compact notation that will be used to describe the states of (1) single grains, (2) sets of grains, and (3) entire packings in this special case.  
\begin{table}
\caption{\label{convention}Compact notation for 2D packings with $Z=4$.}
\begin{ruledtabular}
\begin{tabular}{lccc}
Scale & Number of  & State & Density \\
 & Grains & Variable & of States\\
\hline
Grain & 1 & $g=(w_x, w_y, \theta_1, \ldots, \theta_4)$ & $\rho(g)$\\
Set of Grains & $m$ & $\gamma=(g_1,g_2,\ldots,g_m)$ & $\rho(\gamma)$ \\
Packing & $N$ & $\Gamma=(g_1,g_2, \ldots, g_N \mid \mathcal{C})$ & $\rho(\Gamma)$ \\
\end{tabular}
\end{ruledtabular}
\end{table}
The single-grain state variable is
\begin{equation}
g=(w_x,w_y,\theta_1,\ldots,\theta_4)
\end{equation}
where the Cartesian loads $w_x$ and $w_y$ are the total force borne by a grain in each orthogonal direction and $\theta_i$ are the contact angles.  From these variables, the individual contact forces may be recovered by linear algebra and trigonometry.  This form of $g$ assumes contact number $Z=4$ for every grain.  It may be generalized for grains with $Z \ne 4$ but doing so adds no insight to the physics at this stage.  Note also that certain regions in $g$ describe grain configurations that have one or more negative (tensile) forces.  It will therefore be necessary to restrict the range of $g$ to the non-tensile (stable) region $\mathcal{S}$ when dealing with cohesionless grains, as we do in this paper.

The densities of states in Table~\ref{convention} refer to either single particle or multiple particle states in the corresponding phase space.  For example, the density of single particle states $\rho(g)$ refers to the grains in a single packing or in a single layer of a packing (depending on the context), and it tells how many of those grains exist per unit volume of $g$-space as a function of the location in $g$-space.  Likewise, the density of packing states $\rho(\Gamma)$ refers to a statistical ensemble of granular packings and tells how many of those packings exist per unit volume of $\Gamma$-space as a function of the location in $\Gamma$-space.  The construction list $\mathcal{C}$ contained in $\Gamma$ specifies the exact ordering in which the grains are connected together such that they form a packing.  The density of states $\rho(\gamma)$ is also a multi-particle density of states describing collections of grains with specified single-particle states, but unlike $\rho(\Gamma)$ it does not have a construction list and so it does not specify the location of the grains in physical space nor whether they physically connect to one another in a packing.  Despite the mathematical abstraction, it will be useful in this paper.

\subsection{The Need to Generalize Boltzmann's \textit{Stosszahlansatz}}

\begin{figure}
\includegraphics[angle=0,width=\columnwidth]{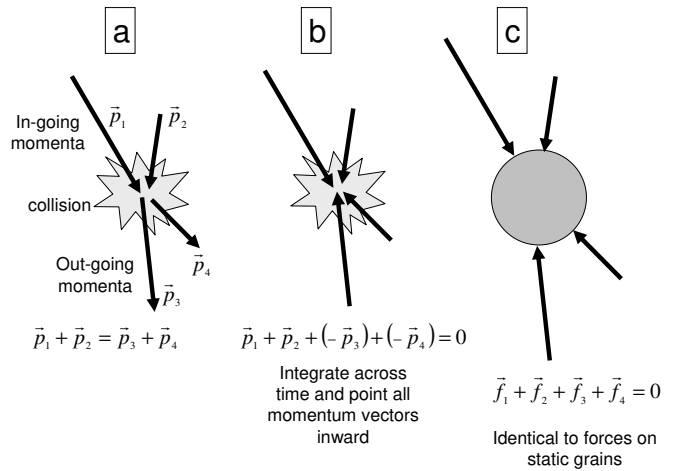}
\caption{\label{collisions} Illustration how momentum vectors in binary collisions of a dilute gas are analogous to contact forces on a static grain.  Boltzmann wrote the \textit{stosszahlansatz} with preference for the in-going momentum vectors, but for static granular materials the symmetries among the contact force vectors must be restored.}
\end{figure}

Boltzmann's \textit{stosszahlansatz}, or his Assumption of Molecular Chaos, is that colliding particles have uncorrelated momenta prior to the collision,
\begin{equation}
F(\bm{p}_1,\bm{p}_2)=f(\bm{p}_1)\cdot f(\bm{p}_2)\label{AMC}
\end{equation}
or more compactly,
\begin{equation}
F_{21}=f_1 f_2
\end{equation}
where $F_{21}$ is the joint probability distribution for the two particles and $f_i$ are the single particle distribution functions.  Are the contact forces that meet together on a grain in a granular packing similarly uncorrelated?  Fig.~\ref{collisions} shows how the relationship of contact forces upon a static grain does bear an analogy to momenta in a gas.
In a particle collision the sum of the momentum vectors is conserved.  Hence, if the outgoing momentum vectors are reversed, then the sum of all four momentum vectors will equal precisely zero.  Putting a little space between the four arrow heads and drawing a circle, we see that this set of four vectors is identical to four contact forces on a static grain.  

However, Eq.~\ref{AMC} assumes non-correlation for the incoming momentum vectors only \cite{reverseAMC}, whereas the idealized granular packings considered here (without gravity) should be more symmetric than this.  If we try to apply Boltzmann's \textit{stosszahlansatz} to granular contact forces, we could group the four forces on the grain into six different pairs, and there is no reason why one of those pairs should be written as uncorrelated to the exclusion of the other five.  Furthermore, Silbert, Grest and Landry have demonstrated through numerical modeling that contact forces meeting together on a grain do have a very strong pattern of correlation and anticorrelation \cite{silbertgrest}.  There is anti-correlation for contacts closer together than roughly $\pi/2$ radians of angular separation, and positive correlation when the angular separation is greater than roughly $\pi/2$.  

Because of these things, it would be incorrect to use the form of Eq.~\ref{AMC} for the granular \textit{stosszahlansatz}.  Instead, we must begin with a statement that is even more fundamental than Eq.~\ref{AMC}, namely, that correlation cannot arise through the closure of loops of forces (or loops of momenta) within the network of these vectors.  This is the {\em generalized} \textit{stosszahlansatz}.  It will be shown below that this generalized statement does indeed reduce to Eq.~\ref{AMC} when causality is assumed to proceed only in the forward time direction.  However, it produces a very different form when ``causality'' (information flow) is assumed to be symmetric in all dimensions of the vector network.  That symmetric form is the one that we need for granular packings.

{\em A priori} arguments have been provided in \cite{metzger2} to justify the generalized \textit{stosszahlansatz}, that correlation does not arise through the closure of force loops in the force vector network.  Appendix A of this paper extends those arguments.  As with thermal systems, the ultimate proof of the \textit{stosszahlansatz} will be its ability to make correct predictions.  If the predictions are correct, then we may claim \textit{a posteriori} that it is indeed the reductionist explanation for the statistics of granular contact forces.

\subsection{Mathematical Form of the Granular \textit{Stosszahlansatz}}

\begin{figure}
\includegraphics[angle=0,width=\columnwidth]{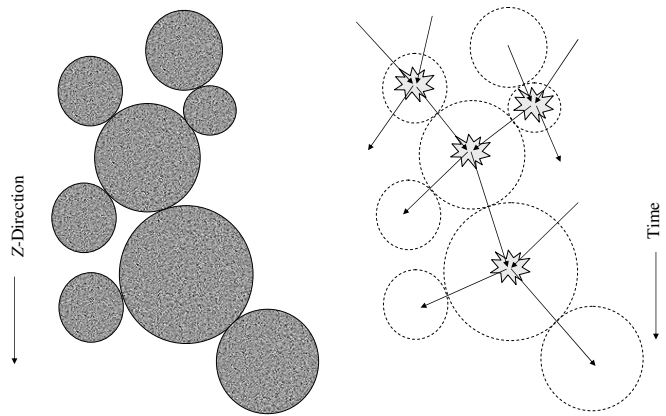}
\caption{\label{networkgraph} Illustration how a network of gas collisions and a network of packed grains may both be represented by network graphs of similar topology.  Identical sets of variables may be used to define a density of states of the vector networks for either system.}
\end{figure}

For a dilute gas it is possible to represent the network of particle collisions occurring through time as a 4-dimensional network graph, with the collisions as the vertices and the particle trajectories as the edges (line segments) connecting the vertices, and with one of the four graphical dimensions representing time.  A lower-dimensional version of this is illustrated in Fig.~\ref{networkgraph}.
It will not be shown here, but the density of states for a block of time in this network can be written using variables that are identical to those used for granular packing densities of states, with the distribution of momenta $f(\bm{p})$ replacing the distribution of contact forces $P(\bm{f})$ and the time dimension replacing any one spatial dimension, and with some differences in the term that defines the allowable sets of momentum vectors (or force vectors) that appear at the collisions (or at the grains).  It is possible to analyze the ensemble of these networks in each case and compare the results.  Here we will show how two different forms of the \textit{stosszahlansatz} are obtained depending only upon the directionality of the information flow through the network.  We will obtain Boltzmann's \textit{stosszahlansatz} when we assume all information travels in one direction across the network (representing time-asymmetry), and we will obtain the granular version of the \textit{stosszahlansatz} when we assume information has propagated across the network symmetrically in all directions.  This will demonstrate that the generalized \textit{stosszahlansatz} really does encompass both versions and reduces to one or the other depending on the assumed direction of information flow.  The density of states we wrote in this section will then no longer be needed in the remainder of the paper.  It will only be used in this section to derive the two mathematical forms of the \textit{stosszahlansatz}.

The multi-particle density of states describing the ensemble of these granular packings (and adaptable to the collision networks of dilute gases) will be written in a form like this,
\begin{eqnarray}
\rho(\Gamma|\{P_j\})&=&\delta(\textrm{Newton's Third Law})\nonumber\\*
 & & \times \Theta(\textrm{No Tensile Forces})\nonumber\\*
 & & \times\Theta(\textrm{Geometric Closure of Loops})\nonumber\\*
 & & \times\delta(\textrm{Sequence $\{P_j\}$})\nonumber\\*
 & & \times\delta(\textrm{Collision Auxiliary})\label{nominaldos}
\end{eqnarray}
For static granular packings of rigid, round, frictionless grains (without the symmetry-breaking effect of gravity), this becomes 
\begin{eqnarray}
\rho(\Gamma|\{P_j(\bm{f})\},P_{4\theta}) & = & \Bigg\{\prod_{(\alpha\epsilon,\beta\delta)\in \mathcal{C}} \delta\!\left(\bm{f}^{\alpha \epsilon}+\bm{f}^{\beta \delta}\right)\Bigg\} \nonumber\\*
& \times & \left\{\prod_{\alpha=1}^{N} \prod_{\epsilon=1}^{4} \Theta\left( f^{\alpha \epsilon} \right)\right\}\ \Theta_\zeta(\Gamma|\mathcal{C}) \nonumber\\*
&\times &\prod_j \delta\!\left(P_j(\bm{f})-\mathcal{P}_j(\bm{f}|\Gamma)\right)\nonumber\\*
 &  \times & \delta( P_{4\theta} -\mathcal{Q}_{4\theta}(\Gamma) )\nonumber\\*
\label{microdos}
\end{eqnarray}
Most of the terms in this equation are just notional and do not have to be defined in enough detail for doing calculations because they will factor out of the equations below and will not affect the results in this paper.  Before describing these terms, we note that Eq.~\ref{microdos} does not have a term to enforce Newton's Second Law on each grain.  Instead, we have defined $g$ and $\Gamma$ such that Newton's Second Law is automatically satisfied for every grain throughout the range of the phase space.  To do this, $g$ was defined with $w_x$ and $w_y$ rather than four contact forces.  This removes two degrees of freedom per grain from the phase space coordinates and thus prevents those combinations of contact forces on a grain that would not have summed to zero.  The individual contact forces may be calculated for any grain by linear algebra and trigonometry from the $w_x$ and $w_y$ with the four contact angles contained in $g$ \cite{N2Lnote}.  Thus, the contact forces appear in several terms of Eq.~\ref{microdos} as functions $\bm{f}^{\alpha \epsilon} = \bm{f}^{\alpha \epsilon}(g_\alpha)$ although the argument was suppressed for compactness.

The first term in Eq.~\ref{microdos} ensures that the density of states is nonzero only in the regions of $\Gamma$ where Newton's Third Law is satisfied for every contacting pair of grains.  The contact force on grain $\alpha$ on its contact $\epsilon$ is $\bm{f}^{\alpha \epsilon}$.  The construction list $\mathcal{C}$ tells which grain/contact $(\alpha\epsilon)$ connects to grain/contact $(\beta\delta)$ and thereby constructs a specific packing from the set of grains specified in $\Gamma$.  This is the most important term affecting the results of this paper.  

For the second term, the Heaviside step function ensures that the density of states is nonzero only where the forces are positive, with positive defined as pointing inwardly on the grains.  This is because we are dealing with cohesionless granular materials (no tensile forces possible).

For the third term, the notional function $\Theta_\zeta(\Gamma|\mathcal{C})$ was introduced by Edwards \cite{edwards1}.  It was defined to evaluate either to zero or unity, being unity only when the locus in $\Gamma$ describes a configuration of grains that form precisely closed loops, with contacting grains just touching each other precisely without any overlapping.  This, along with the constraints on the contact forces, is sufficient to ensure a physically stable granular packing.  $\Theta_\zeta$ has not been defined more specifically because it is not needed in this paper.  

Discussion of the fourth term will be delayed until last due to its importance.  The fifth term is the one that was called the ``collision auxiliary'' in Eq.~\ref{nominaldos}.  This term will take different forms depending on whether we are discussing the case of granular packings (contact force networks) or the case of dilute gases (momentum networks).  The similarities and differences between the two cases are unimportant to the theory developed in this paper, but they are explained in the Appendix if the reader is interested in the analogy between the two cases.   It was called the ``collision auxiliary'' because in the case of dilute gases it would provide more constraints on the set of physically realizable collisions.  In Eq.~\ref{microdos} the term is written in a form relevant only to granular packings.  The delta function ensures that the density of states is nonzero only for packings that have the fabric distribution $P_{4\theta}$ specified within its argument.  This fabric is the joint contact angle distribution $P_{4\theta}(\theta_1,\theta_2,\theta_3,\theta_4)$ providing the probability that all four contacts on an individual grain will take on specified values simultaneously, as discussed in \cite{troadec}.  This joint distribution is important because of the intra-grain correlations between contact forces.  The function $\mathcal{Q}_{4\theta}$ computes this statistical distribution from the set of grains $\{g_\alpha\}$ specified by the locus $\Gamma$.  In the thermodynamic limit with an infinite number of grains, the delta function selects only those packings that have precisely the specified $P_{4\theta}$.  For a practical packing with only a finite number of grains, this term should be defined to allow some statistical fluctuation in $P_{4\theta}$.  However, as stated above, we do not need to define this term so precisely since it is just a notional placeholder and does not affect the rest of the paper.    

The fourth term may be the most confusing, and it factors out of the equations and does not affect the results of this paper, and yet it is the most important term for understanding the purpose of the ensemble.  Why do we begin a paper on a Boltzmann-type transport equation with an ensemble, which is reminiscent of Gibbsian methods?  We do so because we will apply the generalized \textit{stosszahlansatz} to this ensemble and then analyze its statistics to obtain a mathematical form for that \textit{stosszahlansatz}.  That mathematical form can then be used in the Boltzmann-type transport equation and we can abandon the ensemble methods at that time.  But for now, this fourth term defines a set of packings in which the density of single particle states can evolve with respect to translation in one of the spatial dimensions, just as Boltzmann was concerned with dilute gases in which the distribution of momenta would evolve with respect to translation through the time dimension.  Therefore, we subdivide the packing into regions that are subscripted by $j$ so that we can translate through these regions by incrementing $j$.  These regions are defined between successive cross-sectional planes cutting across the packing as shown in Fig.~\ref{conserve} \cite{jnote}.  The delta function ensures that the density of states is nonzero only for packings that have the specified distribution of contact forces $P_j(\bm{f})$ within each region.  At this stage of the paper we do not care what the forms of these specified distributions $P_j$ actually are or how they may evolve.  All we care about is that every packing in the ensemble has the same sequence of distributions $\{P_1(\bm{f}),P_2(\bm{f}),P_3(\bm{f}),\ldots\}$ as all the other packings in the ensemble.  Later this paper will show how the sequence of $P_j$ actually does behave \cite{relaxnote}.  The functions $\mathcal{P}_j(\bm{f}|\Gamma)$ compute the distribution of contact forces from the grains $\{g_\alpha\}$ in each region $j$ specified by the locus $\Gamma$.  As with the similar function $\mathcal{Q}_{4\theta}$ discussed above, for practical packings with only a finite number of grains the term should be defined to allow some statistical fluctuation in the contact force distributions.  However, this term is notional and will factor out from the equations below.  

Finally, we note that there is no term to specify the stress state of the packing.  That is because it is specified implicitly in the $\{P_j\}$.  We can compute the stress in any region as a function of the contact force vectors in that region.  Newton's third law enforced across the packing will ensure that any stresses appearing in adjacent regions are physically possible. 

\begin{figure}
\includegraphics[angle=0,width=\columnwidth]{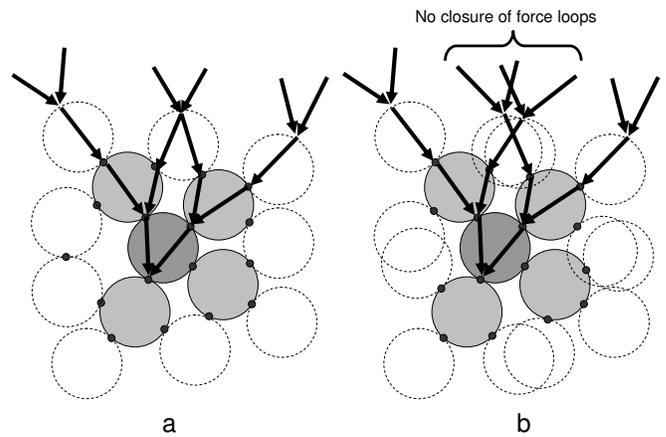}
\caption{\label{noforceloops}(Left) When $\Theta_\zeta$ is kept as part of the density of states, forces form loops and so intersecting forces on the central grain may have pre-correlation, because they have seen parts of each other before.  (Right) When $\Theta_\zeta$ is omitted from the density of states, then force loops do not close and so the forces that meet together on a grain cannot be pre-correlated by the packing.  The only correlation arises from the stability requirements of the central grain, itself.}
\end{figure}

Because the identities of the grains are unimportant, the statistics of the ensemble are unaffected by summing the density of states $\rho$ over any set of permutations of grain exchanges.  At any particular location in $\Gamma$-space, the delta function that enforces Newton's third law will select from those permutations only those that produce a self-consistent packing.  Here for convenience we sum over those permutations that keep the grains within their own regions $j$, denoted by the subset $\{i\}*$ of the set $\{i\}$ of all possible construction lists $\mathcal{C}_i$.  
\begin{equation}
\tilde{\rho}(\Gamma) = \sum_{i\in\{i\}*} \rho(\Gamma)\label{sumrho}
\end{equation}
All terms in Eq.~\ref{microdos} factor out from this sum except those containing $\mathcal{C}_i$ explicitly.  We shall examine this unfactored part in the thermodynamic limit 
\begin{equation}
\Phi= \lim_{\{N_j\}\to\{\infty\}} \sum_{i\in\{i\}*} \Theta_\zeta(\Gamma\mid \mathcal{C}_i)\prod_{\mathcal{C}_i} \delta^2\!\left(\bm{f}^{\alpha\gamma}+\bm{f}^{\beta\delta}\right)\label{phieqn}
\end{equation}
Now we apply the generalized \textit{stosszahlansatz} by removing the closure of force loops in the packings. We do so simply by eliminating the $\Theta_\zeta$ term from Eq.~\ref{microdos}, as explained by Fig.~\ref{noforceloops}.  
This converts the ensemble of closed-loop contact force networks into an ensemble of tree networks as shown in Fig.~\ref{noforceloops}.  If the generalized \textit{stosszahlansatz} is correct, then eliminating $\Theta_\zeta$ will have no effect on the statistics of the contact forces or the single grain states in the ensemble.  

\begin{figure}
\includegraphics[angle=0,width=\columnwidth]{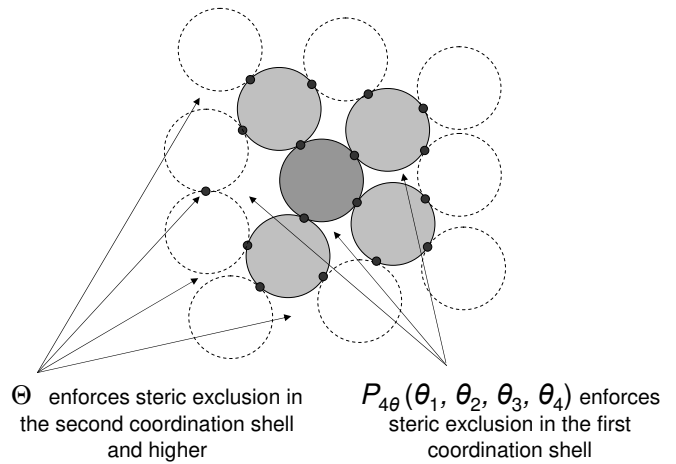}
\caption{\label{stericshells}Illustration of coordination shells.  The central grain is shaded dark.  The first coordination shell is shaded lightly.  The second coordination shell is dashed and unshaded.  Omitting $\Theta_\zeta$ from the density of states removes the closure of force loops but does not affect the fabric of the individual grains.}
\end{figure}

We may note that it was a strategic choice to specify the joint distribution of contact angles $P_{4\theta}$ as the fabric in the fifth term, because it implicitly enforces steric exclusion by evaluating to zero everywhere that adjacent grain contact angles are too close together.  Therefore, while $\Theta_\zeta$ served the purpose of enforcing the non-overlapping of grains in the packing, $P_{4\theta}$ continues to serve this purpose within the first coordination shell of every grain, as illustrated in Fig.~\ref{stericshells}.
Therefore, even though we have eliminated $\Theta_\zeta$, we still have steric exclusion in the first coordination shell and so the set of individual grains in the ensemble will still be valid.  Overlapping of grains will appear only in the second and higher coordination shells \cite{overlapnote}.  

Here it is most important to note that Eq.~\ref{microdos} is not an Edwards ensemble, which would have included all possible sequences $\{P_j\}$.  We do not specify the relative probability of systems having one particular sequence versus any other; we only specify one particular sequence to be retained in this subset of the Edwards ensemble.  The flat measure within this subset of the Edwards ensemble is inherent to Boltzmann's kinetic theory when we assume his \textit{stosszahlansatz}.  That is, in the thermodynamic limit when we drop the closure of loops every tree network becomes identical apart from a trivial relabeling of the branches.  The stronger claims made by Gibbs' and Edwards' flat measures have been avoided in favor of this much weaker claim inherent to Boltzmann's theory.

The analysis of $\Phi$ is shown in the appendix.  Two different results can be obtained depending on the assumed direction that information has propagated through the network.  If we assume that information propagated only in the direction of increasing $j$, then we obtain
\begin{equation}
F_j(\bm{f}_{1},\bm{f}_2)= P_j(\bm{f}_{1}) \cdot P_j(\bm{f}_{2})\label{BAMC}
\end{equation}
which is Boltzmann's form.  We may also derive the symmetric case on the same network, with the information propagating symmetrically in both directions of $j$.  This situation describes, for example, a linearly elastic system of grains after the dynamic stress waves have dissipated and the solution to the Laplace equation remains.  As shown in the appendix, we obtain
\begin{equation}
F_j(g)= [P_j(\bm{f}_{1}) \cdot P_j(\bm{f}_{2}) \cdot  P_j(\bm{f}_{3}) \cdot  P_j(\bm{f}_4)]^{1/2}  \label{symmAMC}
\end{equation}
where $f_i=f_i(g)$ and where the four forces can be located with complete generality around the grain (including 3 contacts on one side of the grain and only 1 contact on the opposite side, etc.).  This is the mathematical form of the granular \textit{stosszahlansatz}.   This form is now symmetric over all four forces, and yet because of the square root it still has dimensions of $[f]^{-2}$.  This is so simple that perhaps we should have guessed the result before actually deriving it.  Clearly we did not need to derive the classical form given in Eq.~\ref{BAMC}, since it is well-known from probability theory that the joint distribution of two independent variables is the product of their individual distributions.  However, the symmetric form given in Eq.~\ref{symmAMC} is unknown in probability theory (to the author's knowledge) because probability theory deals with macroscopic events in time, in which causality occurs asymmetrically in the forward time direction, only.  Eq.~\ref{symmAMC} is the first new result of this paper.

\section{\label{sec:Htheorem} Granular ``H'' Theorem}

\subsection{The Granular Translation Equation}

We shall analyze what happens to the density of layer states defined as in Fig.~\ref{conserve} as its cross sectional line translates in some direction $x$, which is perpendicular to the layer and corresponds to increasing $j$ for the successive $P_j(\bm{f})$.  Therefore, $\rho(g) = \rho_j(g)=\rho(g,x_j)$ although the $j$ and the $x$ shall be suppressed.  

As we translate from $x_j \to x_{j+1}=x_j+\Delta x$ with $\Delta x << D_{\textrm{particle}}$ (the grain diameter), a small fraction of the grains in the layer will no longer be intersected by the line and hence will exit the layer.  Also, some new grains will be intersected by the cross-sectional line and hence they will join the layer.  If the layer contains $M$ grains, then the number of grains expected to leave the layer in $\Delta x$ is $m=M \Delta x/D_{\textrm{particle}}$.  If the fabric is constant across the packing (so that $M$ is constant), then this is also the number of grains joining the layer in $\Delta x$.  For sufficiently small $\Delta x$ the grains exiting the layer will be sufficiently far apart to be statistically independent.  This avoids the need to explicitly account for correlations in the layer.  The probability for a particular set of $m$ grains to leave during $\Delta x$ is therefore
\begin{equation}
P_{\textrm{out}}(\gamma)=\prod_{\alpha=1}^{m} \rho(g_\alpha) \label{leaving}
\end{equation}
The probability for a particular set of $m$ grains to enter during $\Delta x$ can be written in terms of the generalized \textit{stosszahlansatz} from Eq.~\ref{symmAMC}, except that we must be careful because in general $P(\bm{f})$ for contacts that are on one hemisphere of the grains will not be the same as for the other hemisphere because $\rho(g)$ is evolving with $x$ and $P=P[\rho]$.  Therefore, we write the \textit{stosszahlansatz} in the form,
\begin{equation}
\Upsilon^{\pm}(g)=\prod_{\eta^{+}} \surd P'(\bm{f}_{\eta}(g)) \prod_{\eta^{-}} \surd P(\bm{f}_{\eta}(g)) \label{pmAMC}
\end{equation}
where $\eta^{-}$ refers to all the contacts on grain $\alpha$ in the reverse direction of the translation, and $\eta^{+}$ refers to its contacts in the forward direction of the translation.  Likewise, $P$ is the distribution of contact forces in the reverse direction and $P'$ the distribution in the forward direction.  Because the grains entering the layer are statistically independent for sufficiently small $\Delta x$, we may write the probability for a particular set of $m$ grains to enter during that interval as
\begin{equation}
P_{\textrm{in}}(\gamma')=\mathcal{N}\prod_{\alpha=1}^{m} \Upsilon^{\pm}(g'_\alpha) \label{entering}
\end{equation}
where $\mathcal{N}$ is for normalization.  

Because forces are conserved from one layer to the next, the physics are analogous to Boltzmann's binary collisions, except that they are ``$m$-nary'' transitions of grains entering and exiting each layer, and that each iteration $\Delta x$ contains only one of these $m$-nary transitions.  The probability of having a particular transition $\gamma \to \gamma'$ is
\begin{equation}
n(\gamma \to \gamma') = \mathcal{N} \prod_{\alpha=1}^m \rho(g_\alpha) \Upsilon^{\pm}(g'_\alpha)
\end{equation}
The probability that $\gamma$ will go to \textit{any} set of single grain states is
\begin{eqnarray}
n(\gamma \to \textrm{all}) & = & \int_{\mathcal{S}} \mathcal{D}g'_1 \cdots \mathcal{D}g'_m\ \ \mathcal{N} \prod_{\alpha=1}^m \rho(g_\alpha) \Upsilon^{\pm}(g'_\alpha)
\nonumber\\*
\ \label{layertoall}
\end{eqnarray}
where the integrals are carried out only over the stable region $\mathcal{S}$ in $g_i$.  Likewise,
\begin{eqnarray}
n(\textrm{all} \to \gamma) & = & \int_{\mathcal{S}} \mathcal{D}g'_1 \cdots \mathcal{D}g'_m\ \ \mathcal{N} \prod_{\alpha=1}^m \rho(g'_\alpha)  \Upsilon^{\pm}(g_\alpha)\nonumber\\*
& = & \int_{\mathcal{S}} \mathcal{D}g'_1 \cdots \mathcal{D}g'_m\ \ \mathcal{N} \nonumber\\*
& & \times \left( \prod_\alpha \rho(g_\alpha) \Upsilon^{\pm}(g'_\alpha) - \prod_\alpha \rho(g'_\alpha) \Upsilon^{\pm}(g_\alpha)\right)\nonumber\\*
\label{alltolayer}
\end{eqnarray}

To determine the rate of change in $\rho(g)$ during the translation, we write,
\begin{equation}
\frac{\textrm{d}}{\textrm{d}x}{\rho}(g_1) = \int_{\mathcal{S}} \mathcal{D} g_{2} \cdots \mathcal{D}g_{m} \ \left[n(\textrm{all} \to \gamma) - n(\gamma \to \textrm{all})\right]\label{trans2}
\end{equation}
which is the ``translation equation.''

\subsection{Counting States for Granular Packings}

To evaluate how this translation equation behaves we need a functional similar to Boltzmann's $H$.  As a mathematical proof, the $H$ theorem does not demand that $H$ have any physical meaning.  However, for the dilute gas $H$ becomes the negative of Shannon's entropy when the system is in equilibrium, and it is helpful after the mathematical proof is complete to discuss the physical meaning of this.  Following Boltzmann, the granular $H$ shall likewise be (the negative of) a generalization of Shannon's entropy.  We define $H$ so that it will indicate how many packing states correspond to any particular $\rho(g)$.  The ``most entropic'' $\rho(g)$ is the one that arises in the greatest number of packings.  As explained in Ref.~\cite{metzger2}, we may explicitly count the packing states as a functional of $\rho(g)$ except that here it shall be applied to a single layer instead of an entire packing.  First we discretize $\rho(g) \to \nu_{ijklmn}$, where the six arguments of $g$ have been broken into small bins of size $\Delta w$ and $\Delta \theta$ and indexed as $w_{xi},w_{yj},\theta_{1k},\ldots,\theta_{4n}$.  Each bin is further divided into $s$ smaller bins to enable the typical binomial counting without Pauli exclusion,  
\begin{eqnarray}
\Omega\{\nu_{ijklmn}\} & = & \prod_{i}\prod_{j}\prod_{k}\prod_{l}\prod_{m}\prod_{n}\left[\frac{(s-1+\nu_{i\ldots n})!}{(s-1)!\ (\nu_{i\ldots n})!}\right]\nonumber\\*
& & \times \left( \sum_{i\ldots n} \nu_{i\ldots n}\right)! \ \left[\Upsilon_{i\ldots n}\Psi_{i\ldots n}\right]^{\nu_{i\ldots n}} \label{counting}
\end{eqnarray}
See \cite{metzger2} for the details.  $s$ shall drop out of the equations in the continuum and thermodynamic limits when $\Delta g =(\Delta w)^2 (\Delta \theta)^4 \to 0$ as $N \to \infty$. $\Psi$ evaluates to zero when any of the $g$ states imply tensile forces, and it evaluates to unity otherwise.  For simplicity we will restrict all further mathematics to the stable region $\mathcal{S}$ so we may drop $\Psi$ from the expressions.

We define $H$ as
\begin{equation}
H \triangleq  - \lim_{\begin{array}{c}N\to \infty\\ \Delta g \to 0\end{array}}\log\Omega
\end{equation}
The natural logarithm of $\Omega$, using Stirling's approximation, may be written
\begin{eqnarray}
\log\Omega & = \sum_{i,j,k,l,m,n} & \left[(s-1+\nu_{i\ldots n})\log(s-1+\nu_{i\ldots n})\right.\nonumber\\*
& & -(s-1)\log(s-1)-\nu_{i\ldots n}\log\nu_{i\ldots n}\nonumber\\*
& & \left.+ \nu_{i\ldots n}\log\Upsilon_{i\ldots n}\right]\label{counting1}
\end{eqnarray}
Expanding the first term in a Taylor series around $\nu_{i\ldots n}=0$, setting $s=N \Delta g$, and taking the continuum and thermodynamic limits such that $s>>\nu_{i \ldots n}$, we obtain
\begin{equation}
H = \int_{\mathcal{S}} \mathcal{D}g\ \rho(g) \log\frac{\rho(g)}{\Upsilon(g)}
\end{equation}
The functional $H=H[\rho(g)]$ is a measure of the number of states in $\rho(\gamma)$ that correspond to $\rho(g)$, and is in fact (the negative of) the generalization of Shannon's entropy for granular contact forces.  In the translation, $\rho(g)$ must be allowed to evolve layer-by-layer, and therefore so must $P(\bm{f})$.  Hence, we distinguish between the hemispheres of a grain and use the form of $\Upsilon$ from Eq.~\ref{pmAMC} to write,
\begin{equation}
H = \int_{\mathcal{S}} \mathcal{D}g\ \rho(g) \log\frac{\rho(g)}{\Upsilon^{\pm}(g)} \label{Hdef}
\end{equation}

\subsection{Behavior of $\rho$ and $H$ in the Translation}

With this metric in hand, the question we wish to address is whether $(\textrm{d}/\textrm{d}x)H \le 0$ during the translation described above.  Differentiating $H$, 
\begin{eqnarray}
\frac{\textrm{d}}{\textrm{d}x}H  =  -\int_{\mathcal{S}} \mathcal{D}g\ \left\{\frac{\textrm{d}}{\textrm{d}x}\rho(g) \left[ 1 + \log \frac{\Upsilon^{\pm}(g)}{\rho(g)} \right]\right.\nonumber\\*
\left.+\rho(g)\frac{\textrm{d}}{\textrm{d}x}\log\Upsilon(g)\right\}
\end{eqnarray}
The last term (call it $\chi$) may be expanded,
\begin{eqnarray}
\chi & = & -\int_{\mathcal{S}} \mathcal{D}g\ \rho(g)\frac{\textrm{d}}{\textrm{d}x}\log\Upsilon(g)  \nonumber\\*
 & = &  -\int_{\mathcal{S}} \mathcal{D}g\ \rho(g)\frac{\textrm{d}}{\textrm{d}x}\log\prod_\eta \surd P(\bm{f}_\eta(g))\nonumber\\*
 & = &  -\frac{1}{2}\sum_\eta \int_{\mathcal{S}} \mathcal{D}g\ \rho(g)\frac{\textrm{d}}{\textrm{d}x}\log P(\bm{f}_\eta(g))
\end{eqnarray}
and then evaluated by changing the integration from a sum over all grains to a sum over all contacts,
\begin{eqnarray}
\chi & =  & - \frac{1}{2}\int_0^\infty \textrm{d}\bm{f}\ P(\bm{f}) \frac{\textrm{d}}{\textrm{d}x}\log P(\bm{f})\nonumber\\*
 & = & - \frac{1}{2}\frac{\textrm{d}}{\textrm{d}x} \int_0^\infty \textrm{d}\bm{f}\ P(\bm{f})\nonumber\\*
 & = & 0
\end{eqnarray}
Substituting Eqs.~\ref{layertoall}, \ref{alltolayer}, and~\ref{trans2} into $(\textrm{d}/\textrm{d}x)H$,
\begin{eqnarray}
\frac{\textrm{d}}{\textrm{d}x}H & = & \mathcal{N} \int_{\mathcal{S}}  \mathcal{D}g_1 \cdots\mathcal{D}g_m \int_{\mathcal{S}} \mathcal{D}g'_1 \ldots \mathcal{D}g'_m \nonumber\\*
& & \times \left( \prod_\alpha \rho(g_\alpha) \Upsilon^{\pm}(g'_\alpha) - \prod_\alpha \rho(g'_\alpha) \Upsilon^{\pm}(g_\alpha)\right) \nonumber\\*
& & \times\ \left[ 1 + \log \frac{ \Upsilon^{\pm}(g_1)}{\rho(g_1)} \right]
\end{eqnarray}
Since we integrate over $\mathcal{S}$ for all $g_\alpha$ and $g'_\alpha$, we may swap variables and average the various equivalent expressions to obtain,
\begin{eqnarray}
\frac{\textrm{d}}{\textrm{d}x}H & = & \frac{\mathcal{N}}{2m} \int_{\mathcal{S}}  \mathcal{D}g_1 \cdots\mathcal{D}g_m \int_{\mathcal{S}} \mathcal{D}g'_1 \ldots \mathcal{D}g'_m \nonumber\\*
& & \times \left( \prod_\alpha \rho(g_\alpha) \Upsilon^{\pm}(g'_\alpha) - \prod_\alpha \rho(g'_\alpha) \Upsilon^{\pm}(g_\alpha)\right) \nonumber\\*
& & \times\ \log \frac{ \prod_\alpha \rho(g'_{\alpha}) \Upsilon^{\pm}(g_\alpha)}{\prod_{\alpha} \rho(g_\alpha)\Upsilon^{\pm}(g'_\alpha)} 
\end{eqnarray}
By inspection of the integrand we see that it is never positive for any part of the region of integration.  Hence, 
\begin{equation}
\frac{\textrm{d}}{\textrm{d}x}H \le 0
\end{equation}
Furthermore, $(\textrm{d}/\textrm{d}x)H=0$ if and only if
\begin{equation}
\prod_\alpha \rho(g'_\alpha) \Upsilon^{\pm}(g_\alpha) = \prod_\alpha \rho(g_\alpha) \Upsilon^{\pm}(g'_\alpha)\ \  \forall\ \  g_\alpha,g'_\alpha \label{equilib}
\end{equation}
and this is true if and only if $(\textrm{d}/\textrm{d}x)\rho(g)=0$ for all $g$.  When that is the case, then $P=P'$.  This proves that the bulk of the packing must exist in a relaxed state.  More will be said about this state, below.

Furthermore, as shown in the next section, Eq.~\ref{equilib} defines a sufficient condition to solve $\rho(g)$ and so by Eq.~\ref{Hdef} it is also sufficient to evaluate $H$.  This produces the smallest possible value of $H$, since for any other value of $H$ $(\textrm{d}/\textrm{d}x)H \ne 0$.  Since this derivation is valid in every possible orientation of the layer (because the derivation was general and because the \textit{stosszahlansatz} is symmetric), then it must also be valid in the direction of decreasing $x$,
\begin{equation}
\frac{\textrm{d}}{\textrm{d}(-x)}H \le 0 
\end{equation}
This is not true for the dilute gas, because Boltzmann's {\em stosszahlansatz} is not symmetric \cite{reverseAMC}.  But for the granular case with the greater symmetry,
\begin{eqnarray}
\frac{\textrm{d}}{\textrm{d}x}H \le 0 & \textrm{and} & \frac{\textrm{d}}{\textrm{d}x}H \ge 0 
\end{eqnarray}
which implies
\begin{equation}
\frac{\textrm{d}}{\textrm{d}x}H \equiv 0
\end{equation}
must be true always, for the bulk of an infinitely large, homogeneous granular packing in the special case considered here. No exploration of phase space is required for the packing to relax.  Relaxation is assured through spatial relationships, not temporal ones.  

The reason this theory is limited to the bulk of the packing, rather than predicting relaxation of stresses moving away from the boundary of a container, is that there are two things that can cause $H$ to change:  evolving stresses, and evolving fabric.  At the present we do not know how to predict the evolution of fabric in the boundary layer, and so it is impossible to separate out the effect of the relaxation of stresses in that same region.  By assuming that the fabric is constant as in the bulk of a very large packing, this theory has concluded that the spatial distribution of stresses is such that it will minimize $H$ within that fabric.  

Now we shall consider the physical meaning of $H$.  By its definition, minimum $H$ corresponds to maximum contact force entropy.  Thus, the maximum entropy condition has been proven.  This proof depends only on the \textit{stosszahlansatz}, which shall be validated by comparing its predictions against numerical data.  The conclusion is that Edwards' hypothesis is not necessary to assert maximum contact force entropy.

\section{A New Derivation of the Density of States and $P(f)$ Without Edwards' Hypothesis}

The foregoing granular $H$-theorem tells us that for infinitely large packings maximum entropy $(\textrm{d}/\textrm{d}x)H \equiv 0$ persists in every layer, and by extension to very large, finite packings it persists in the majority of layers away from the boundaries but with greater fluctuations occurring in smaller packings.  The sufficient and necessary condition for maximum entropy is given by Eq.~\ref{equilib}.  This can be written in the form,
\begin{equation}
\prod_{\alpha=1}^m \frac{\rho(g_\alpha)}{\Upsilon^{\pm}(g_\alpha)} = \prod_{\alpha=1}^m \frac{\rho(g'_\alpha)}{\Upsilon^{\pm}(g'_\alpha) }\ \  \forall\ \  g_\alpha,g'_\alpha
\end{equation}
which implies that either side of the equation may be written as equal to a constant.  Taking the logarithm,
\begin{equation}
\sum_{\alpha=1}^m \log\frac{\rho(g_\alpha)}{\Upsilon^{\pm}(g_\alpha)} = C \label{cons1}
\end{equation}
Since this is valid for every possible ``$m$-nary'' transition of grains during a translation distance of $\Delta x$, its most general solution is when $C$ is written as a linear combination of all quantities that are conserved in the translation.  Since we did not specify the orientation or the direction of the translation, this result is valid for all.  As discussed in Sec.IIA, the total force in every orientation must be conserved.  Fabric is also conserved by definition of the problem.  Therefore, 
\begin{equation}
\log\frac{\rho(g_\alpha)}{\Upsilon(g_\alpha)} = \beta_x w_x^\alpha + \beta_y w_y^\alpha + \mu(\theta_1,\ldots,\theta_4) \label{cons2}
\end{equation}
for every $g_\alpha \in \mathcal{S}$.  The first two terms are for force conservation and the last is for fabric conservation.  (We can write the first two terms using stress tensor notation if we wish, and hence explicitly include shear stresses.)  Rearranging to obtain $\rho$,
\begin{equation}
\rho(g) = G(\theta_1,\ldots,\theta_4) \Psi(g) \Upsilon(g) e^{-\beta_x w_x -\beta_y w_y}\label{conneqn}
\end{equation}
where we recall that $\Psi$ is the function that enforces the bounds on $\mathcal{S}$, evaluating either to unity or zero if the cohesionless grain is stable or unstable, respectively.  $G(\theta_1,\ldots,\theta_4)$ is the fabric partition factor.  Eq.~\ref{conneqn} is identical to the equation derived in \cite{metzger2} by Gibbs' {\em most probable distribution} method.  That method {\em assumed} the maximum entropy condition as a subset of the claims in Edwards' hypothesis, but here it has been derived.  Remembering that $\Upsilon$ is a functional of $P(\bm{f})$ by Eq.~\ref{pmAMC} and that
\begin{equation}
P(\bm{f}) = \int_\mathcal{S} \mathcal{D}g\ \  \rho(g)\ \delta^2(\bm{f} - \bm{f}(g)) \label{colleq}
\end{equation}
we see that Eqs.~\ref{conneqn} and~\ref{colleq} form a recursion so that $\rho$ may be solved numerically when stress and fabric are specified.  Solving for $\rho(g)$ provides everything that can be known about the single particle density of states, including $P(\bm{f})$, so it makes testable predictions.  

\section{Empirical Validation of the Granular \textit{Stosszahlansatz}}

As with Boltzmann's theory, the translation equation and the granular $H$ theorem depend upon the validity of the \textit{stosszahlansatz}.  It can be tested only by comparison with experimental or simulation data.  First we must obtain predictions from the theory so we have something to compare.  The numerical solution of $\rho(g)$ from the recursion equation has been obtained (in approximation, for the isotropic case) and the details of the method are provided in \cite{metzger2}.  The method uses Monte Carol integration, randomly sampling the region of integration with a flat measure.  This is performed in a recursion, with $\rho(g)$ obtained from $P(f)$ and vice-versa until convergence.  This recursion begins from an arbitrary initial condition, for example $P(f)=\delta(f-1)$.  Several different initial conditions were checked and all converged to the same result.  This method was used to produce a representation of $\rho(g)$ consisting of 12 billion grain configurations, which was sufficient for very smooth statistics.  On an average desktop workstation the algorithm converged for a sample of one million grains in about one minute, with a larger number of grains taking proportionately longer.  Several different algorithms and different approximations to the math were used and they resulted in only minor variations in the statistical results, demonstrating the robustness of the basic form of the solution.  Since a packing's fabric $P_{4\theta}(\theta_1,\ldots,\theta_4)$ will evolve as the packing is sheared, it is desirable to force the numerical solution to a $\rho(g)$ that has some particular fabric that is found in a real case.  Thus, it is necessary to weight the Monte Carlo sampling to include some classes of contact angle configurations more often than others.  The weighting may be determined iteratively by adjusting the weight factor used in the algorithm until the desired fabric is obtained.  This weight factor, multiplied by zero in the regions of steric exclusion, is in fact the fabric partition factor $G(\theta_1,\ldots,\theta_4)$ that appears in Eq.~\ref{conneqn}.  Similarly, the overall stress in the solution $\rho(g)$ may be driven to any desired state by weighting the Monte Carlo sampling with the Boltzman factor shown in Eq.~\ref{conneqn}.

\begin{figure}
\includegraphics[angle=0, width=\columnwidth]{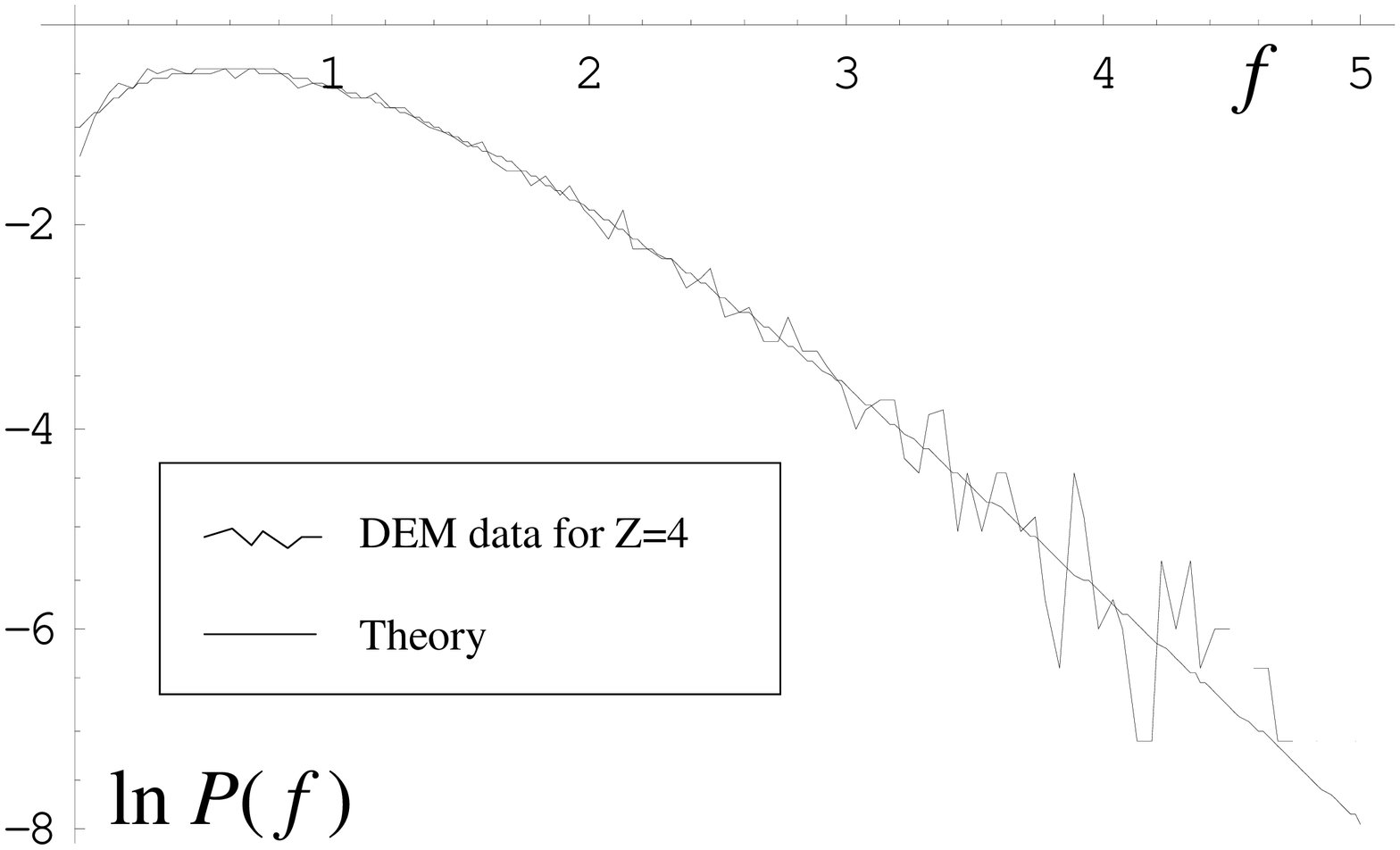}
\caption{\label{z4compare}Semi logarithmic $P\left(f\right)$ obtained in numerical solution of the translation equation (theory) and from the $Z=4$ population of a DEM simulation.  The statistical fluctuations in the tail of the DEM data (amplified on the log axis), do not exist in the theory because of the thermodynamic limit.}
\end{figure}

This numerical solution has been compared to numerical data from discrete element modeling with idealizations approaching those of the theory, and a subset of the results have appeared in a Letter \cite{metzger3} with an archival-length paper to follow.  The simulation used 17,000 grains that were 2D, round, frictionless, and cohesionless.  The fabric used in the theory implies that they are monodisperse (in that the steric exclusion angle was assumed to be precisely $\pi/3$ radians).  However, the simulation used a small polydispersity of 1.5 to avoid crystallization of the grains.  To approach the isotropic idealization the grains were deposited into a rectangular, rigid-walled container randomly and without gravity, and then expanded in diameter while allowed to push each other around until they jammed.  The grains within four grain diameters of the walls were discarded from the statistical analysis to avoid boundary effects, since the theory describes the bulk of infinitely large packings (where boundaries do not exist).  Residual kinetic energy was viscously dissipated until the forces had negligible dynamical fluctuation, representing the idealization of a perfectly static packing.  The grains had a linear spring contact law, but the packing was kept as close as could be achieved near the limit of jamming so that minimal compression of the contacts occurred.  This approximates the perfect grain rigidity and the isostaticity of the theory wherein the exact form of the contact law becomes irrelevant.  The data show that with these idealizations the predictions of the theory are validated.  An example of the correspondence between theory and empirical results is shown in Fig.~\ref{z4compare},
which shows $P(f)$ obtained from the theory and from the DEM data for the $Z=4$ grains.  Discussion of the results of $Z\ne 4$ grains is found in \cite{metzger3} and will be expanded in the future archival-length paper.  It should be noted that there are no free parameters in the theory that could have been adjusted to obtain the correspondence; the good agreement occurs automatically.  Thus, the \textit{stosszahlansatz} has been validated along with the maximum entropy condition that it predicts for this isotropic case.  We may conclude that the reductionist explanation for granular contact force statistics is that they exist at maximum entropy, and that is because correlations cannot arise through the closure of loops of grains.  This statement should be further tested in non-isotropic and less idealized cases.

\section{Summary and Conclusions}

By following Boltzmann's method but with symmetric information flow in the collision network, this paper has shown that the layer-by-layer relationships in granular contact forces, like the time-evolution of momenta in a dilute gas, must cause their statistical distribution to relax to the form that represents maximum entropy.  This paper also derived a version of Boltzmann's \textit{stosszahlansatz} and a version of Shannon's entropy (negative $H$) relevant to granular contact forces, and found a way to derive $\rho(g)$ without recourse to Edwards' hypothesis.  It has been demonstrated previously with more results to appear in a future archival paper that the predictions of $\rho(g)$ are in outstanding agreement with numerical simulations.  

Boltzmann's $H$ theorem for the dilute gas tells us that every tiny segment of the Poinc\'are cycle is dominated by the Maxwellian distribution, so that only a tiny amount of dynamics is needed to depart from any non-Maxwellian state and settle into maximum entropy.  Both the Poinc\'are ergodicity and the Boltzmannian relaxation are thus attained by traveling along the trajectory in phase space.  Static granular packings, on the other hand, do not travel along any trajectory in phase space.  The granular translation theorem developed above depends upon contact forces maintaining {\em static} spatial relationships, not traveling and interacting through time.  Whereas Boltzmann's proof obviated the need to travel the {\em entire} Poinc\'are cycle and showed that the system need travel only a tiny segment of it to justify the assumptions of statistical mechanics, the granular $H$ theorem shows that a granular packing need not travel through $\Gamma$ space {\em at all}, because it has special relationships between its own layer-wise subspaces built into its single locus in $\Gamma$ space.  The author proposes that this special feature of granular packings deserves a name, and ``self-ergodic'' seems appropriate.  That is, the system must exist in a relaxed state by nature of its internal relationships (self-enforced and affecting itself), and this produces the same statistical characteristics that an ergodic theorem seeks to establish for kinetic systems exploring all conserved-energy states with equal probability.  Thus the terminology:  self + ergodic.

There are of course many significant classes of granular packings that will not be relaxed to maximum entropy, but these have not been discussed here.  These include the boundary regions near the container walls of a packing, as well as granular packings that were prepared to have abrupt changes in fabric somewhere within their bulk so that the density of states cannot be in its most relaxed state within the narrow band of grains on both sides of the interface.  The difficulty in applying this theory to boundary regions is that the fabric does not remain constant near the boundary, whereas the theory assumes that the fabric is constant.  Another case not described by the theory is the thin layer of grains at the top surface of a packing in gravity, where the self-weight introduces non-negligible stress gradients.  However, solution of the more symmetric case considered here opens the door to solving more complex problems.  

This theory has been developed only for the case of rigid, round grains without friction.  This avoids the complication of rotational degrees of freedom for the grains.  The author does not know how to define the grain states to include torques and rotational degrees of freedom so that they would have sufficient symmetry for the analysis.  The assumptions of the theory also exclude cases with highly compressed grains, which are far from isostaticity \cite{makse}.

The theory has been developed assuming $Z=4$ for every grain, although frictionless 2D packings with low polydispersity are known to have grains with $Z=3$, $Z=5$ and occasionally $Z=6$.  It seems possible to generalize the theory to allow for different values of $Z$ as long as the average is still 4 for isostaticity.  In this paper, the grain state $g$ was defined to have just the two variables with units of force, $w_x$ and $w_y$, plus the four contact angles.  We may alternatively use the two variables $t=w_x+w_y$ and $s=(w_x-w_y)/t$.  So, for grains with $z=3$ we would define $g_3=\{t,\theta_1,\theta_2,\theta_3\}$, and this would be sufficient to define the grain's entire state allowing us to solve all the contact force values on the grain by linear algebra and trigonometry.  For $Z=4$ we would use $g_4=\{t,s \cdot t,\theta_1,\ldots,\theta_4\}$, which is equivalent to the treatment given in this paper.  Then, for $Z=5$ we need one additional state variable in $g_5$ to allow us to solve all five contact forces on the grain.  This additional variable could simply be the value of one of the contact forces, but to maintain symmetry among the contacts it would presumably be better to define something analogous to a quadrupole term and thus continue the sequence of $t$ and $st$, which represent the monopole and dipole terms, respectively.  Obtaining a numerical solution to the resulting theory would be a simple extension of the existing $Z=4$ algorithm.

It is unknown whether this theory will work for an ordered (crystalline) granular packing.  Experimental and numerical studies have shown that even the microscopic variations in packing geometry of crystalline packings are sufficient to break the symmetry and relax $P(f)$ to a typical form \cite{crystalrelax}.  This theory may take advantage of this by keeping translation distances $\Delta x$ small enough that the microscopic packing variations are orders-of-magnitude larger.  Thus even with a crystalline packing the assumptions behind Eqs.~\ref{leaving} and~\ref{entering} should be valid and the theory may work.  Nevertheless, it is possible that correlations will arise in the regular pattern of closed loops, violating the generalized \textit{stosszahlansatz}.  It would be interesting to study how much disorder is required for the theory to work.

One application of this theory in particular was hypothesized by the author several years ago \cite{arxiv}.  The idea is that, since granular contact forces are in a state of equilibrium at maximum entropy subject to the layer-by-layer conservation of forces in each direction, then it is possible to define a rank-2 tensor ``temperature'' for the contact forces and even a coordination number ``chemical potential'' to explain the partition of stress and fabric fluctuations throughout a granular packing.  This approach may lead to a full theory of rheology.  These observations were apparent when numerical solutions to the theory first proved robust and convergent, even before this formal proof of maximum entropy was accomplished.  That is because the numerical convergence discussed in the earlier publications was in fact an empirical demonstration of the same concept.  Once the ``self-ergodic'' nature of granular materials is identified, then the temperature and entropy concepts fall out rather straightforwardly. A future publication will be forthcoming to explain these concepts along with a series of discrete element modeling simulations that have been performed to test them and to draw further conclusions.

\begin{appendix}

\section{\textit{A Priori} Arguments for the Granular \textit{Stosszahlansatz}}

\begin{figure}
\includegraphics[angle=0,width=\columnwidth]{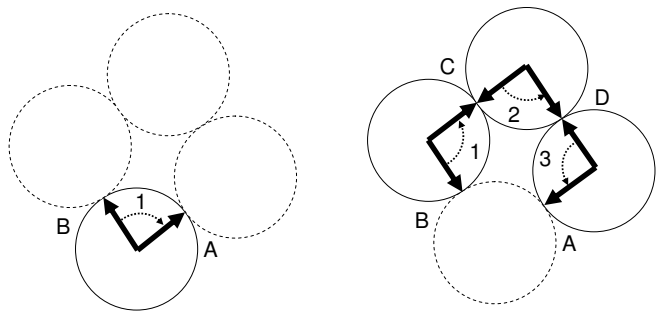}
\caption{\label{loopcorr} Correlations between neighboring contact forces may arise two different ways.  (Left) They arise through intra-grain stability requirements, resulting in a strong, two-point correlation between the forces at $B$ and $A$.  (Right)  They arise through a series of intra-grain stability requirements working grain-by-grain around the loops, resulting in weak higher-order correlations.  The example here shows a four-point correlation through the loop, $B$ to $C$ to $D$ to $A$, which will be much weaker than the direct two-point correlation from $B$ to $A$.}
\end{figure}

As illustrated in Fig.~\ref{loopcorr},
correlations between neighboring contacts on the same grain arise either through the grain itself or through the loops in the packing.  A typical loop is four or more grains, so going the long way around a loop induces a relatively weak four-point correlation, but going the short way between the two contacts (staying intra-grain) induces a much stronger two-point correlation.  

\begin{figure}
\includegraphics[angle=0,width=0.30\textwidth]{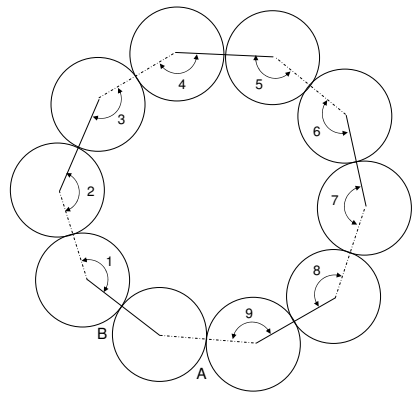}
\caption{\label{longloop} Contacts that are highly correlated are close to $\pi$ radians apart on the same grain \cite{silbertgrest}.  Hence, a closed loop composed of highly correlated pairs of contacts can turn only very slowly and must pass through a very large number of grains.}
\end{figure}

Furthermore, we may note that small loops composed of $N$ grains, $3 \le N < 8$, are formed through adjacent pairs of contacts that are closer to the non-correlating $\pi/2$ angle than the correlating $\pi$ radians of separation (see \cite{silbertgrest}), and hence there should be minimal correlation in each two-point leg of a force loop.  The composite correlation going all the way around the loop must therefore be exceedingly weak.  On the other hand, force loops that pass through a larger number of grains and hence closer to $\pi$ radians of separation for each two-point leg of the loop will require a vastly larger $N$ to slowly turn through the full $2 \pi$ radians to close the loop as illustrated in Fig.~\ref{longloop}.
Because $N$ is so large ($N>8$), we would therefore expect these $N$-point correlations to be very weak, as well.  

The total correlation between a pair of contacts on a grain must be the sum of information from \textit{all} the loops in the packing that contain the grain in question.  Due to the disorder of the packing and the large number of loops that contain the same grain, some correlating and some anti-correlating, it is expected that the contributions from increasingly larger loops of grains will be increasingly decoherent and largely cancel one another.  

From these arguments there is good reason to assume \textit{a priori} that the intra-grain contribution to the correlations is the dominant one and that closed-loop contributions may be discarded.  This will be the granular \textit{stosszahlansatz}.  The ultimate proof of this {\em stosszahlansatz} is its ability to make valid predictions.  Comparison with empirical data \cite{metzger4, metzger3} have validated its predictions.  

\section{Two Forms of the Collision Auxiliary in the Density of States}

The purpose of the ``collision auxiliary'' term in Eq.~\ref{nominaldos} is to restrict the allowable set of vectors ``colliding'' (meeting together) at the nodes (grains or particle collisions).  Already the vector sums of the forces or momenta are automatically conserved (\`a la Newton's Second Law) by the proper selection of phase space coordinates.  However, in both granular mechanics and in the kinetic theory of dilute gases there are additional restrictions imposed by physics upon these sets of vectors.  

For the case of the dilute gas, in addition to conservation of momentum there is the requirement for the conservation of energy, and even that is not sufficient to define the allowable sets of vectors appearing at each collision.  A pair of colliding momentum vectors may meet at any arbitrary magnitudes and angle relative to one another.  The magnitudes and directions of the two outgoing momentum vectors are then determined by the precise form of the interaction potential between the particles.  Thus, the collision auxiliary term in Eq.~\ref{nominaldos} must be written to specify the set of allowable outgoing vectors based upon the incoming vectors and the interaction potential.  

For the case of granular packings, on the other hand, the sets of allowable force vectors appearing on each grain are not so precisely constrained.  In addition to the conservation of force from one hemisphere to the next (Newton's Second Law) there is only the requirements of steric exclusion.  There is no analog of interaction potential or conservation of energy to determine two of the vectors as a function of the other two.  Precisely because of this additional freedom, granular packings exhibit memory in their fabric, the statistical preponderance of their contact angles resulting from past disturbances of the packing.  Therefore, an ensemble of granular packings would need to have the current state of its fabric specified in order to be a completely defined ensemble.  The ``collision auxiliary'' is therefore used for this purpose.  

Although the collision auxiliary plays two very different roles in granular versus kinetic ensembles, they have been lumped together here under the name ``collision auxiliary'' because they both play the role of defining the sets of vectors appearing at the nodes insofar as required by the physics.  The important point to note is that it does not matter which form we use in this paper, either the kinetic or the granular form, because this term factors out from the sum in Eq.~\ref{sumrho} and does not affect the results.  Thus, the results we obtain are valid for either granular or kinetic ensembles and we can compare the mathematical form derived for the \textit{stosszahlansatz} in one case with the mathematical form derived in the other.

\section{Derivation of the \textit{Stosszahlansatz} for Two Cases of Information Flow}

Beginning from Eq.~\ref{phieqn}, we expand the product over $C_i$,
\begin{eqnarray}
\Phi &=& \lim_{\{N_j\}\to\{\infty\}} \prod_j \sum_{i_j=1}^{N_j!} \prod_{\alpha_j=1}^{N_j}  \prod_{\gamma=1}^2 \nonumber\\*
& & \times\prod_{\beta_j=\alpha_j}^{N_j} \prod_{\delta=3}^4 \delta_{\mathcal{C}_i}\ \delta^2\!\left(\bm{f}^{\alpha\gamma}+\bm{f}^{\beta\delta}\right)\label{asymmstep}
\end{eqnarray}
where the Kronecker delta function $\delta_{\mathcal{C}_i}$ is nonzero only for contact pairs contained within list $\mathcal{C}_i$.  We have ordered the contact pairs in $\mathcal{C}_i$ so that the grain in the direction of lower $j$ is $\alpha$ and contacts in the increasing $j$ hemisphere are numbered 1 and 2.  Note that the product in $\beta$ is only from $\alpha$ to $N$, the upper triangle in the $(\alpha,\beta)$ plane, since Newton's third law is enforced only once for every pair, and this produces the correct number of delta functions so that the units of $\tilde{\rho}$ are correct.  In writing it this way, we have preferentially given it an asymmetry in the $j$ direction.

Now we must make the critical assumption about causality in this network.  First we attempt to recover Boltzmann's \textit{stosszahlansatz} for the dilute gas.  We do so by treating the information asymmetrically in the $j$ direction, so that the probability of finding a certain set of grains in $j$ depends only on the set of grains in $j-1$.  Therefore, for this case we may start from Eq.~\ref{asymmstep}.  Commuting the sum in $i$ with the first two products,
\begin{eqnarray}
\Phi & = &\lim_{\{N_j\}\to\{\infty\}} \mathcal{M} \prod_j \prod_{\alpha_j=1}^{N_j}  \prod_{\gamma=1}^2 \sum_{i_j=1}^{N_j!} \nonumber\\*
& & \times \prod_{\beta_j=\alpha_j}^{N_j} \prod_{\delta=3}^4 \delta_{\mathcal{C}_i}\ \delta^2\!\left(\bm{f}^{\alpha\gamma}+\bm{f}^{\beta\delta}\right)\nonumber\\*
& = & \lim_{\{N_j\}\to\{\infty\}} \mathcal{M} \prod_j \prod_{\alpha_j=1}^{N_j}  \prod_{\gamma=1}^2 \sum_{i_j=1}^{N_j!} \nonumber\\*
& & \times \sum_{\beta_j=\alpha_j}^{N_j} \sum_{\delta=3}^4 \ \delta^2\!\left(\bm{f}^{\alpha\gamma}+\bm{f}^{\beta\delta}\right)
\end{eqnarray}
This was renormalized by $\mathcal{M}$ since each of the terms in the product is now a sum of many delta functions instead of just one delta function.  This also introduced a small error inside the limit in that the product of the sums expands to a sum of the product of all possible combinations of the delta functions with replacement, whereas it should have been without replacement.  Taking the limit, the cross-terms with non-representative distributions of single grain states $\rho_g$ become a set of zero measure and so the error vanishes.  Next, we expand the delta function 
\begin{eqnarray}
\Phi & = & \lim_{\{N_j\}\to\{\infty\}} \mathcal{M} \prod_j \prod_{\alpha_j=1}^{N_j}  \prod_{\gamma=1}^2 \sum_{i_j=1}^{N_j!} \nonumber\\*
& & \times \sum_{\beta_j=\alpha_j}^{N_j} \sum_{\delta=3}^4 \ \lim_{\Delta f \to 0} \frac{1}{\Delta f} \lim_{\Delta \theta \to 0} \frac{1}{\Delta \theta} \nonumber\\*
& & \times  \left\{ \Theta\left[f^{\beta\delta}-f_k(f^{\alpha\gamma})\right]-\Theta\left[f^{\beta\delta} - f_{k+1}(f^{\alpha\gamma})\right]\right\}\nonumber\\*
& & \times \left\{ \Theta\left[|\theta^{\beta\delta}-\theta_l( \theta^{\alpha\gamma})|-\pi\right]\right.\nonumber\\*
& & \ \ \ \ \ \ \ \ \ \ \ \ \ \ -\left.\Theta\left[|\theta^{\beta\delta}-\theta_{l+1}(\theta^{\alpha\gamma})|-\pi\right]\right\}
\label{phi2b}
\end{eqnarray}
where the subscripts $k$ and $l$ define the $k^{\textrm{th}}$ interval of the force axis as the interval containing $f^{\alpha \gamma}$ and the $l^{\textrm{th}}$ interval of the angle axis as the interval containing $\theta^{\alpha\gamma}$.  Taking the summations,
\begin{eqnarray} 
\Phi &=& \mathcal{M} \prod_j \lim_{\{N_j\}\to\{\infty\}} \lim_{\Delta f \to 0} \lim_{\Delta \theta \to 0} \frac{1}{\Delta F} \ \frac{1}{\Delta \theta} \prod_{\alpha=1}^N  \prod_{\gamma=1}^4 n_{kl}^{(j)}\nonumber\\*
\ 
\end{eqnarray}
where $n_{kl}^{(j)}$ is the number of contacts in region $j$ that fall into the $k^{\textrm{th}}$ and $l^{\textrm{th}}$ bins.  Taking the limits converts $n_{kl}^{(j)} \to P_{j}(\bm{f})$ 
\begin{equation}
\Phi =  \mathcal{M} \prod_j \prod_{\alpha=1}^{N_j} P_j(\bm{f}^{\alpha 1}) P_j(\bm{f}^{\alpha 2}) 
\end{equation}
This says that the probability of finding a member of the ensemble with a particular set of force pairs intersecting on a common grain in the $j^{\textrm{th}}$ layer is $\prod_{\alpha=1}^{N_j} F_j(\bm{f}^{\alpha 1},\bm{f}^{\alpha 2})$, where the distribution of intersecting forces in the $j^{\textrm{th}}$ layer of the ensemble is,
\begin{equation}
F_j(\bm{f}_{1},\bm{f}_2)= P_j(\bm{f}_{1}) P_j(\bm{f}_{2}) \label{symmresult}
\end{equation}
We have derived Boltzmann's \textit{stosszahlansatz}.  We obtained it from a statement about the topology of the network---no closure of force (or momentum) loops---rather than simply writing it as a statement borrowed from probability theory.  

Now we can put into the analysis the symmetry that is correct for a granular packing, so that causality operates equally in all directions.  We do so by modifying Eq.~\ref{asymmstep} by including all contacts on each grain and by including the entire $(\alpha,\beta)$ plane, so that every contact is considered twice:  once looking in the $+j$ direction and again looking in the $-j$ direction.  (We can also relax the restriction that contacts 1 and 2 must be on a particular hemisphere of the grain with contacts 3 and 4 on the opposite hemisphere.  This more symmetric treatment allows grains to have three contacts on one side of the grain and only one contact on the opposite side.)  But now that we are multiplying over the entire $(\alpha,\beta)$ plane there are twice as many delta functions and so in effect $\Phi$ has  been squared,
\begin{eqnarray}
\Phi^2 &=& \lim_{\{N_j\}\to\{\infty\}} \prod_j \sum_{i_j=1}^{N_j!} \prod_{\alpha_j=1}^{N_j}  \prod_{\gamma=1}^4 \nonumber\\*
& & \times\prod_{\beta_j=1}^{N_j} \prod_{\delta=1}^4 \delta_{\mathcal{C}_i}\ \delta^2\!\left(\bm{f}^{\alpha\gamma}+\bm{f}^{\beta\delta}\right)\label{symmstep}
\end{eqnarray}
The analysis is now symmetric in $j$ and proceeds identically as before, with the end result,
\begin{equation}
\Phi^2 =  \mathcal{M} \prod_j \prod_{\alpha=1}^{N_j} P_j(\bm{f}^{\alpha 1}) P_j(\bm{f}^{\alpha 2}) P_j(\bm{f}^{\alpha 3}) P_j(\bm{f}^{\alpha 4})
\end{equation}
so that by taking the square root of both sides we conclude
\begin{equation}
F_j(w_x, w_y, \theta_1, \ldots, \theta_4)= [P_j(\bm{f}_{1})\ P_j(\bm{f}_{2})\ P_j(\bm{f}_{3})\ P_j(\bm{f}_{4})]^{1/2}\label{granAMC}
\end{equation}
This is the granular \textit{stosszahlansatz}.  Both this version and Eq.~\ref{symmresult} say that the topology of the network does not pre-correlate the intersecting force vectors.

\end{appendix}


\begin{thebibliography}{99}
\bibitem{reverseAMC} It can easily be shown that the outgoing vectors are correlated whenever $f(\bm{p},t)$ is not the Maxwell-Boltzmann distribution.
\bibitem{N2Lnote} Similarly, if we apply this method to an ensemble of dilute gas collision networks (in four space+time dimensions), then the phase space variables would not include the individual momentum vectors $\bm{p}_1$ and $\bm{p}_2$ for each collision, nor the outgoing momentum vectors $\bm{p}'_1$ and $\bm{p}'_2$.  Instead, it would consist of the sum of the incoming momentum vectors at each collision, $\bm{P}$, along with the colliding momentum vector angles prior to the collision, $(\theta_1, \phi_1)$ and $(\theta_2, \phi_2)$ and after the collision $(\theta'_1, \phi'_1)$ and $(\theta'_2, \phi'_2)$.  The magnitudes of the incoming and outgoing momentum vectors could then be recovered by linear algebra and trigonometry.  Note that this differs from the granular case only in that there is one fewer conserved component of momentum than there are dimensions in the collision network:  there are $P_x$, $P_y$ and $P_z$, but no $P_t$ corresponding to the time dimension.  Despite this difference, the method works identically for both the granular and dilute gas cases.
\bibitem{jnote}  In the case of the dilute gas where the ensemble describes collision networks, these regions are layers of width $\Delta t$ in the time dimension.
\bibitem{relaxnote}  In particular, if we leave out the fourth term from Eq.~\ref{microdos} so that the ensemble reverts to an Edwards ensemble, then in the thermodynamic limit essentially all of the packings in the ensemble will demonstrate a certain kind of relaxation in $\{P_j\}$.  Any packings that have a different sequence $\{P_j\}$ without this relaxation will thus comprise a set of zero measure in the Edwards ensemble.
\end{thebibliography}
\end{document}